\documentclass[10pt,letterpaper]{article}
\usepackage{opex3}
\usepackage{amsmath}
\usepackage{units}
\usepackage{cite}
\usepackage{epstopdf}
\usepackage{epsfig}
\usepackage{mathbbol}
\usepackage[normalem]{ulem}	
\usepackage{color}

\begin{document}

\title{Detection of Bessel beams with digital axicons}

\author{Abderrahmen Trichili,$^{1}$ Thandeka Mhlanga,$^{2,3}$ Yaseera Ismail,$^{3}$ Filippus S. Roux,$^{2}$ Melanie McLaren,$^2$ Mourad Zghal,$^{1}$ and Andrew Forbes,$^{2,3}$}
\address{$^{1}$University of Carthage, Engineering School of Communication of Tunis (Sup'Com), GreS'Com Laboratory, Ghazala Technopark, 2083, Ariana, Tunisia\\$^{2}$CSIR National Laser Centre, P.O. Box 395, Pretoria 0001, South Africa\\$^{3}$School of Physics, University of KwaZulu-Natal, Private Bag X54001, Durban 4000, South Africa}
\email{aforbes1@csir.co.za} 

\begin{abstract}
We propose a simple method for the detection of Bessel beams with arbitrary radial and azimuthal indices, and then demonstrate it in an all-digital setup with a spatial light modulator.  We confirm that the fidelity of the detection method is very high, with modal cross-talk below 5\%, even for high orbital angular momentum carrying fields with long propagation ranges. To illustrate the versatility of the approach we use it to observe the modal spectrum changes during the self-reconstruction process of Bessel beams after encountering an obstruction, as well as to characterize modal distortions of Bessel beams propagating through atmospheric turbulence.
\end{abstract}



\section{Introduction}
Since their discovery in 1987 by Durnin\cite{Durnin1,Durnin2}, Bessel beams have been extensively studied due to their nominally non-diffracting behaviour and their ability to self-reconstruct after encountering an obstruction \cite{twist, mazilu1,sorter2} .  These beams are characterized by a radial wave vector ($k_{r}$) and azimuthal index ($\ell$), which results from their helical wave front structure. As a result Bessel beams carry orbital angular momentum (OAM), even down to the single photon level\cite{obstruction3,photon1,photon2}. However an ideal Bessel beam requires an infinite amount of energy; this beam is practically approximated in a finite region by Bessel Gaussian (BG) Beams\cite{BGbeam}. Such beams have been generated using annular ring-slits in the far field\cite{Durnin2,Durnin3}, axicons in the near field\cite{axicon1,axicon2}, as well as the digital equivalent of both\cite{SLM1,SLM2,SLM3,SLM4}. These beams have been further explored by generating their superpositions\cite{superposition}, and converting them into vector BG beams\cite{BG}. An emerging area of research is optical communication with the spatial modes of light, where Bessel beams are also mooted to play a role, yet very little work has been done on the topic of two-dimensional detection of such modes \cite{sorter,mazilu2,mazilu3}.   

In this paper we demonstrate the detection of Bessel beams by a simple scheme comprising only a helical axicon and a lens. We outline the concept, illustrate how it may be implemented optically and then demonstrate it with digitally encoded phase-only holograms. We apply the tool to the self-healing process of Bessel beams after an obstruction as well as to Bessel beams propagating through turbulence, and observe the changing radial and azimuthal spectrums for the first time. Our results will be relevant to future studies in optical communication with Bessel beams. Such fields are interesting for communication purposes since they carry OAM over extended distances in a nominally non-diffracting manner, and hence may be advantageous for signal delivery to distance receivers.
\section{Theoretical background}
\label{theory}
\subsection{Bessel-Gaussian modes}

The Bessel-Gaussian (BG) modes \cite{BGbeam} in polar coordinates, are given by
\begin{equation}
E_{\ell}^{BG}(r,\Phi,z)=\sqrt{\frac{2}{\pi}}J_{\ell}\left( \frac{z_{R}k_{r}r}{z_{R}-iz} \right)\exp ( i\ell\Phi-ik_{z}z)\exp \left
(\frac{ik_{r}^{2}z w_{0}^{2}-2kr^{2}}{4(z_{R}-iz)} \right),
\label{bgmode}
\end{equation}
where $\ell$ is the azimuthal index (a signed integer), ${\rm J}_{\ell}(\cdot)$ is the Bessel function of order $\ell$; $ k_{r}$ and $k_{z}$ are the radial and longitudinal wave numbers. The initial radius of the Gaussian profile is $w_{0}$ and the Rayleigh range is $z_{R}=\pi w_{0}^{2}/\lambda$. The propagation constant $k$ and the parameters $k_{r}$ and $k_{z}$ are related by $k^{2} = k_{r}^{2} + k_{z}^{2}$.  While BG modes are nominally non-diffracting, they nevertheless have a finite propagation distance when generated in the laboratory, given by

\begin{equation}
z_{max} = \frac{w_{0}\lambda}{2\pi k_{r}}.
\label{zmax}
\end{equation}

Bessel beams also exhibit reconstruction of the amplitude and phase of the beam after encountering an obstruction \cite{obstruction,obstruction2}.
For such beams, there is a minimum distance behind an obstacle of radius $R_{obs}$ before reconstruction occurs. This distance represents the shadow region which is given by
\begin{equation}
z_{min}=\frac{2\pi R_{obs}}{k_{r}\lambda}.
\end{equation}

The BG modes form a complete orthonormal basis in terms of which an arbitrary paraxial laser beam may be expanded. In the case of Bessel beams we note that there are two indices used to describe the field: the discrete parameter, $\ell$, and the continuous parameter $k_r$.  The former determines the helicity of the wavefronts and is related to the OAM content of the field, while the latter determines the spacing of the intensity rings observed in Bessel beams.

\subsection{Concept}
\begin{figure}[htbp]
\centerline{\includegraphics[width=12cm]{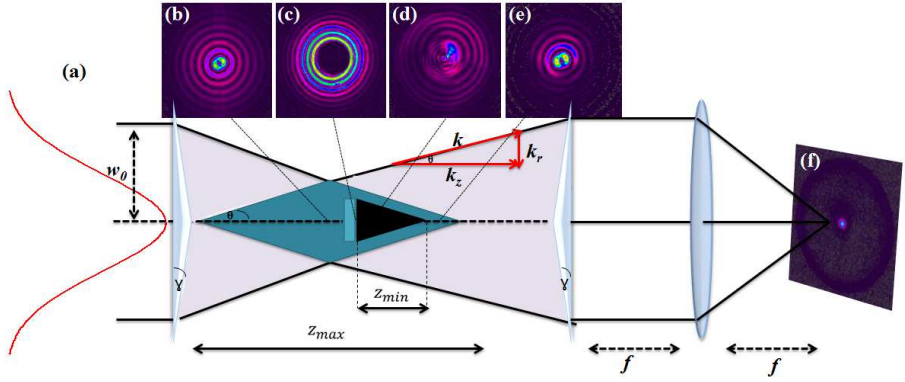}}
\caption{A diagram illustrating the generation and the detection of Bessel-Gaussian beams. (a) The BG beam is generated using a programmed hologram of an axicon, illuminating by a Gaussian beam, and exists in a finite region, $z_{max}$. An obstacle placed in the center of the BG region obstructed the generated beam for a minimum distance, $z_{min}$, after which the BG mode reconstructs. (b-e) experimental beam images of a Bessel beam of order $\ell=1$ at four different positions. (f) The BG beam is detected at the far field of a programmed hologram of a second axicon.} 
\label{concept}
\end{figure}

The task is to find the modal content of the field for all values of $\ell$ and $k_r$, which we will show can be achieved with a simple optical set-up comprising a lens and a digital hologram encoded to represent an axicon.  Recall that a Gaussian beam illuminating an axicon produces a BG beam as the output.  From the reciprocity of light the reverse process must convert a BG beam back into a Gaussian beam.  Herein lies the possibility of detecting particular BG modes, since Gaussian modes may readily be detected by single mode fibers. The concept is shown schematically in Fig.~\ref{concept}.  Consider first a ray-based analysis following a heuristic argument: an incoming Gaussian mode is converted by the first axicon to a BG mode of radial wavevector $k_r = k (n-1) \gamma$, where $n$ is the refractive index of the axicon and $\gamma$ is the axicon cone angle.  This results in conical refraction at an angle $\theta = (n-1)\gamma = k_r/k$. If this BG mode passes through an identical (reversed) axicon, then the refracted rays are collimated, or equivalently, the Gaussian mode is generated again.  If, on the other hand, the cone angle of the second axicon does not match the cone angle of the incoming BG beam, then the outgoing rays will not be perfectly corrected and equivalently a pure Gaussian mode will not be formed. This detection is therefore $k_r$ specific and is reminiscent of a conventional lens telescope but with conical axicons rather than spherical lens.  With the addition of a spiral plate with transmission function $\exp(i\ell \phi)$ the detection method becomes specific to the BG order $\ell$ as well.

\begin{figure}[htbp]
\centerline{\includegraphics[width=10cm]{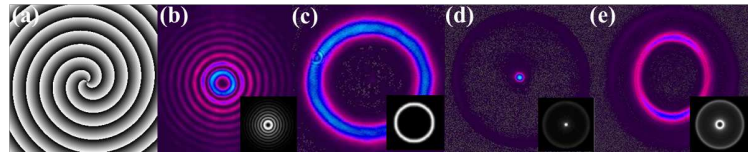}}
\caption{Experimental images of (a) a digital hologram for the detector of a BG mode with $\ell= 3$ and (b) a BG mode profile of $\ell= 3$ and (c) its Fourier transform (annular ring).  The signal at the detector is shown for the scenarios of (d) matching $k_r$ and $\ell$ and (e) matching in $\ell$ but no matching in $k_r$. The black and white insets show the theoretical results.}
\label{spekbess}
\end{figure}

This heuristic argument made more concrete by considering the problem from a physical optics perspective and employing digital holograms for the detection.  The detection hologram may be written as
\begin{equation}
t_{SLM} = \exp(i \tilde{k}_r r - i\ell \phi),
\label{axtr}
\end{equation}

\noindent where the first term represents an axicon to detect a BG with a radial wavevector of $\tilde{k}_r$ and the second term specifies the order, $\ell$. Such a hologram is shown in Fig.~\ref{spekbess} (a) and the BG mode that it will detect in Figs.~\ref{spekbess} (b) and (c).  An inner product measurement is performed optically with the same set-up by considering the signal at the origin of the focal plane of the lens \cite{OAM1}.  The resulting signal can be calculated numerically from
\begin{equation}
g_{out} = {\cal F} \left\{ E_{\ell}^{BG} \right\} \otimes {\cal F} \left\{ t_{SLM} \right\} ,
\label{axconv}
\end{equation}
where $g_{out}$ represents the field at the output plane (focal plane of the lens), $\cal F$ is the Fourier transform, $\otimes$ denotes the convolution process and $E_{\ell}^{BG}$ is the incoming BG beam defined in (\ref{bgmode}). The angular spectrum of a BG mode and the Fourier transform of the transmission function both have the shape of an annular ring. Provided that the radii of these annular rings (which represent the $k_{r}$ values of the modes) are equal, the convolution of these rings will produce a bright spot with a Gaussian profile in the center of the output plane, as shown in Fig.~\ref{spekbess} (d). This central peak is surrounded by a ring of twice the radius.
If there is a mismatch in the respective radii ($ k_{r}$ values) the central spot will itself become a small ring with a low intensity in the center, which will cause a negligible signal on the detector, as shown in Fig.~\ref{spekbess} (e). To quantify this we note that the width of the annular ring (${\cal F} \left\{ E_{\ell}^{BG} \right\}$) is governed by the radius of the Gaussian envelope of the BG mode. On the other hand, the width of the ring due to ${\cal F} \left\{ t_{SLM} \right\}$ is determined by the size of the SLM and is therefore much smaller than the corresponding width for the BG mode. We'll therefore assume that the ring for the axicon transmission function is vanishingly thin. The convolution of the two rings produces a function consisting of two rings with radii that are respectively equal to the sum and difference of the radii of the original rings. Thus if the original radii were equal the convolution produces a central spot. Conversely, if these original radii differ the intensity at the center of the output is given by $\exp[-(\Delta R/w_0)^2]$, where $\Delta r$ is the difference between the original radii. For $\Delta R > 1.5 w_0$ the intensity at the center is essentially zero and the corresponding functions are considered to be orthogonal. Likewise, if the $\ell$ value of the BG mode is different from that of the transmission function of the SLM, they won't canceled during the convolution process. Such a mismatch in $\ell$ values will cause the central peak in the convolution to have a phase singularity in the center and thus a central intensity null, which will produce a negligible signal on the detector. Hence the BG mode detection method is sensitive to both radial ($k_{r}$) and azimuthal ($\ell$) indices.
\section{Experimental Setup and Results}
\begin{figure}[htbp]
\centerline{\includegraphics[width=10cm]{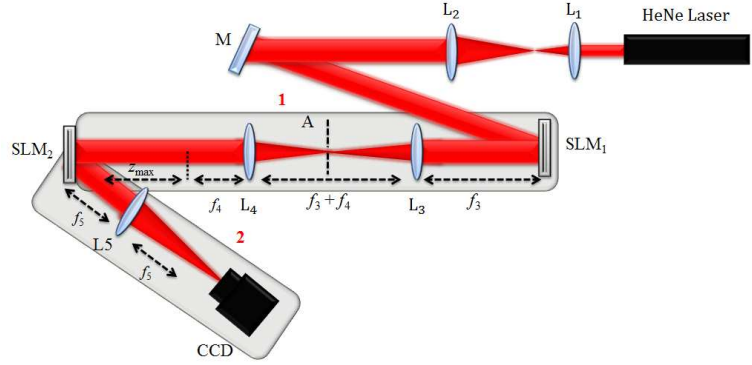}}
\caption{A schematic of the experimental setup for accomplishing the decomposition of a Bessel field. The Lenses $\mathrm{L_{1}}$, $\mathrm{L_{2}}$, $\mathrm{L_{3}}$, $\mathrm{L_{4}}$ and $\mathrm{L_{5}}$ have focal lengths $f_{1}=100$ mm, $f_{2}=300$ mm, $f_{3}=500$ mm, $f_{4}$ =$500$ mm and $f_{5}$=150 mm, respectively. A is the filtering aperture. $\mathrm{SLM_{1}}$ and $\mathrm{SLM_{2}}$ denote the two spatial light modulators and M represents a mirror.  The detector was a CCD camera.}
\label{setup}
\end{figure}

The experimental realization of the BG mode decomposition comprises of two parts: (1) the generation of a BG beam with known parameters (modal profile) and (2) the detection of this beam by modal analysis.  This is accomplished by the optical system shown in Fig.~\ref{setup}, where the created BG beam on $\mathrm{SLM_{1}}$ is assumed to be our ``unknown'' beam. A HeNe laser was expanded with a 3$\times$ telescope and directed onto a spatial light modulator (SLM), denoted as $\mathrm{SLM_{1}}$, with a beam width of $w_0 = 1$ mm. The SLM (Holoeye, PLUTO-VIS, $1920 \times 1080$ pixels, with a pixel pitch of $8 \mu$m) was calibrated for a $2\pi$ phase shift at a wavelength of 633 nm. $\mathrm{SLM_{1}}$ was programmed with the conical phase of an axicon, plus a helical phase with an azimuthal index $\ell$ ranging from -10 to 10. 

The resulting image was filtered through the $4f$ imaging system, and propagated a distance of $z_{max} = 340$ mm (for $k_r = 31250$ rad/m) onto the detection SLM, denoted as $\mathrm{SLM_{2}}$, where the transmission function was scanned through the spectrum of $\ell$ and $k_r$ values and the resulting signal detected by a CCD camera placed after a Fourier transforming lens ($\mathrm{L_5}$). 

\begin{figure}[htbp]
\centerline{\includegraphics[width=10cm]{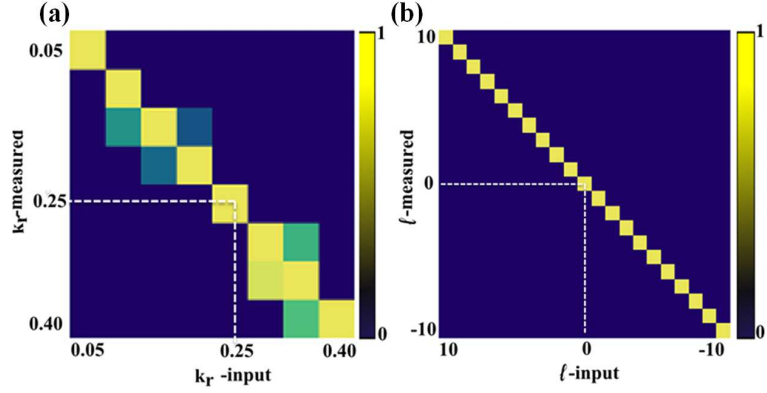}}
\caption{(a) Bessel beam radial, $ k_{r}$, decomposition for $\ell=1$. Units of $k_r$ are rad/pixel. (b) Bessel beam azimuthal, $\ell$, decomposition for $ k_{r}=0.25$ rad/pixel.}
\label{graphs}
\end{figure}


A full modal decomposition was done in $\displaystyle k_{r}$ and $\ell$ at the plane $\displaystyle z = z_{max}$ with the results shown in Fig.~\ref{graphs} (a) and Fig.~\ref{graphs} (b). The uncertainty in detection of the order $\ell$ is clearly negligible while that for the radial wavevector is approximately 5\% (one std dev).  It is clear that a wide range of Bessel modes can be detected quickly and accurately with this scheme.  Next, we illustrate the versatility of our approach by applying it to two perturbation studies: the self-healing of Bessel beams after an obstacle and the propagation of Bessel beams through turbulence.  We use our detection method to experimentally observe the change in modal spectrum during these processes.

\subsection{Bessel Reconstruction}
\begin{figure}[htbp]
\centerline{\includegraphics[width=10cm]{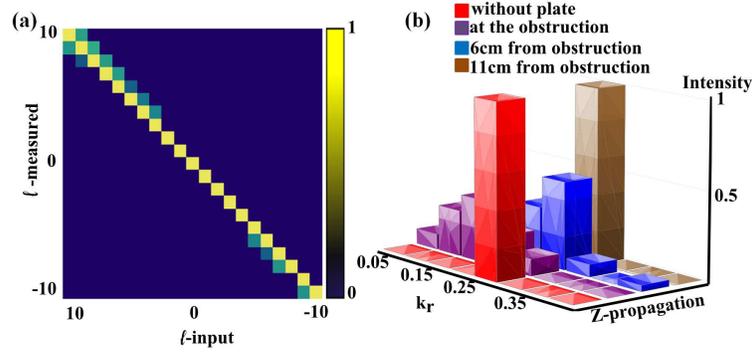}}
\caption{(a) Azimuthal decomposition ($\ell$ detection) of the fully obstructed beam and (b) $k_{r}$ decomposition without an obstruction and then at three planes with the obstruction.}
\label{withoutobstruction}
\end{figure}

A circular opague disk, with a radius of $R_{obs}=300$ $\mu$m, was used as the obstruction.  The disk was initially placed at $\frac{3}{4} z_{max}$ for a BG of $ k_{r}=0.25$. The detection was done at $z = \frac{3}{4} z_{max}$ while the disk was moved away from the detection plane until exceeding the self-healing distance of $z_{min}=9.5$ cm. The radial and azimuthal spectrum was measured before the obstruction and then at various distances after the obstruction until the self-healing process completed.  We observed (see Fig.~\ref{withoutobstruction}) minimal azimuthal distortion of the mode due to the obstruction, but significant broadening of the radial modes.  This broadening reduces as the beam self-heals, returning to the initial spectrum after the self-healing distance.  While the self-healing of Bessel beams has been studied extensively before, this is the first time that the process has been observed using modal analysis.
 
\subsection{Bessel propagation through turbulence}
\begin{figure}[htbp]
\centerline{\includegraphics[width=10cm]{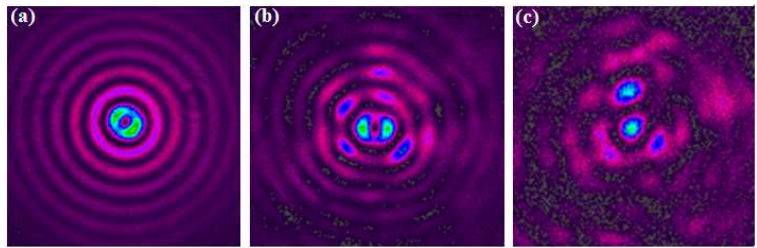}}
\caption{Images of a Bessel-Gaussian mode profile for $\ell= 1$ (a) without turbulence, after passing a turbulence of (b) SR=0.2 and (c) SR=0.03.}
\label{turbulence1}
\end{figure}

\begin{figure}
\centerline{\includegraphics[width=10cm]{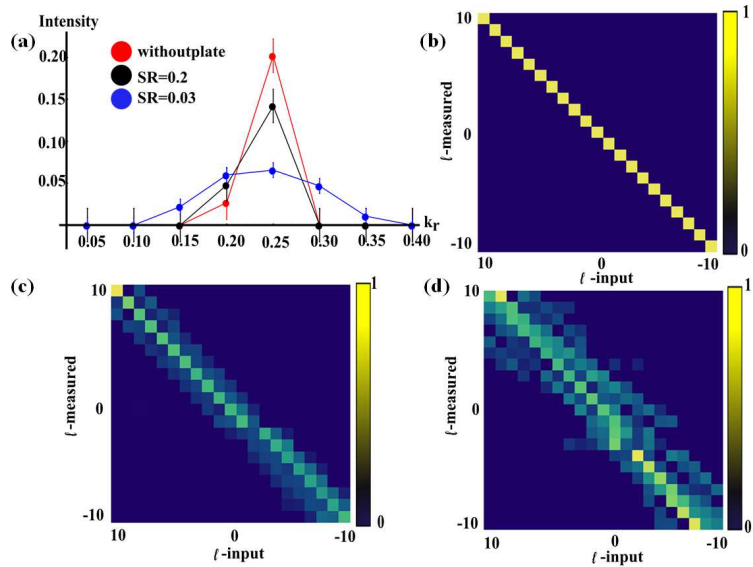}}
\caption{ (a) $k_{r}=0.25$ rad/pixel decomposition for different strehl ratio's. (b) $\ell$ decomposition spectrum without turbulence. (c) and (d) $\ell$ decomposition spectrum for SR=0.2 and SR=0.03, respectively.}
\label{turbulence2}
\end{figure}

Finally, we applied our tool to the study of Bessel beams propagating through turbulence, a topic that has received much theoretical attention of late.  We simulated atmospheric turbulence using a diffractive plate encoded for Kolmogorov turbulence, which for the purposes of this study we characterize by the Strehl ratio \cite{SR}.  The turbulence plate was placed at $\frac{1}{2} z_{max}$ and the detector at $z = z_{max}$.  Two turbulence strengths were used corresponding to Strehl ratios of SR = 0.2 and SR = 0.03, with the impact on the Bessel modes shown in Fig.~\ref{turbulence1}.  Without the plate the results are identical to those shown earlier: narrow $k_r$ and $\ell$ spectrums with little cross-talk, as seen in Figs.~\ref{turbulence2} (a) and (b).  At medium turbulence levels (SR = 0.2), the $k_r$ spectrum broadens and so does the OAM spectrum [Figs.~\ref{turbulence2} (a) and (c)], becoming wider [Figs.~\ref{turbulence2} (a) and (d)] as the turbulence becomes very strong (SR = 0.03).  These results are consistent with that predicted by theory\cite{turbulence1,turbulence2}, and serves to illustrate the versatility of the tool. 

\section{Conclusion}
We have presented a versatile technique to experimentally realize the detection of Bessel beams using digital axicons programmed on a spatial light modulator. We have shown the ability to distinguish both the radial and azimuthal indices of such beams, a core requirement for optical communication protocols if the bit rate per photon is to be increased by exploiting all the degrees of freedom of spatial modes.  In addition we have considered two applications of the tool and observed the modal changes to an incoming Bessel beam due to both amplitude and phase perturbations resulting from an opaque obstacle and a turbulence plate, respectively. The ability to modally resolve such fields will find uses in both quantum and classical studies. 

\section{Acknowledgment}
This  work  has  been supported  by African Laser Centre (ALC)  project ``Towards spatial mode control in fibres for high bit rate optical communication''.  The authors wish to thank Angela Dudley for useful advice.

\begin{thebibliography}{99}
\bibitem{Durnin1}J. Durnin,``Exact solutions for nondiffracting beams. I. The scalar theory,'' \josaa {\bf4}(4), 651-654 (1987).
\bibitem{Durnin2}J. Durnin, J. J. Miceli, and J. H. Eberly, ``Diffraction-Free Beams,'' \prl {\bf58}(15), 1499-1501 (1987).
\bibitem{twist}D. McGloin, and K. Dholakai, ``Bessel beams: diffraction in new light,''  Contemp. Phys.{\bf46}(1),15-28 (2005).
\bibitem{mazilu1} M. Mazilu, D. J. Stevenson, F. Gunn-Moore, and K. Dholakia  ``Light beats the spread: non-diffracting beams,'' Laser Photon. Rev. {\bf4}(4), 529-47 (2010).
\bibitem{sorter2} A.Dudley, M.Lavery, M. Padgett, A. Forbes, ``Unraveling Bessel Beams," Opt. Photon. News, {\bf24}(6), 22-29 (2013)
\bibitem{obstruction3}M. McLaren, T. Mhlanga, M. J. Padgett, F. S. Roux, and A. Forbes, ``Self-healing of quantum entanglement after an obstruction,'' Nat. Commun. {\bf5}:3248 (2014).

\bibitem{photon1}M. McLaren, M. Agnew, J. Leach, F. S. Roux, M. J. Padgett, R. W. Boyd, and A. Forbes, ``Entangled Bessel-Gaussian beams, '' \opex {\bf20}(21), 23589-23597 (2012).
\bibitem{photon2}H. C. Ramírez, R. R. Alarc\'on, F. J. Morelos, P. A. Q. Su, J. C. G. Vega, and A. B. U'Ren, ``Observation of non-diffracting behavior at the single-photon level, '' \opex {\bf20}(28), 29761-29768 (2012).
\bibitem{BGbeam}F. Gori and G. Guattari, ``Bessel-Gauss beams,'' \oc {\bf64}(6), 491-495 (1987).
\bibitem{Durnin3}M. A. Mahmoud, M. Y. Shalaby, and D. Khalil, ``Propagation of Bessel beams generated using finite-width Durnin ring,'' \ao {\bf52}(2), 256-263 (2013).
\bibitem{axicon1}R. M. Herman and T. A. Wiggins, ``Production and uses of diffractionless beams,''\josaa {\bf8}(6), 932-942 (1991).
\bibitem{axicon2}J. Alrt and K. Dholakia, ``Generation of high-order bessel beams by use of an axicon,'' \oc {\bf177}, 277-301 (2000).
\bibitem{SLM1}A. Vasara, J. Turunen, and A. T. Friberg, ``Realization of general nondiffracting beams with computer-generated holograms,''\josaa {\bf6}(11), 1748-1754 (1989).
\bibitem{SLM2}C. Paterson, R. Smith, ``Higher-order Bessel waves produced by axicon-type computer-generated holograms,'' \oc {\bf124}, 121-130 (1996).
\bibitem{SLM3}R. Bowman, N. Muller, X. Zambrana-Puyalto, O. Jedrikiewicz, P. Di Tramapani, and M.J. Padgett, ``Efficient generation of Bessel beam arrays by means of an SLM,'' Eur. Phys. J. Special Topics {\bf199}, 159-166 (2011).
\bibitem{SLM4}Z. Y. Rong, Y. J. Han, S. Z. Wang, and Cheng-Shan Guo, ``Generation of arbitrary vector beams with cascaded liquid crystal spatial light modulators,'' \opex {\bf22}(2), 1636-1644 (2014).
\bibitem{superposition}R. Vasilyeu, A. Dudley, N. Khilo, and A. Forbes, ``Generating superpositions of higher-order Bessel beams," \opex {\bf17}(26), 23389-23395 (2009).
\bibitem{BG}A. Dudley, Y. Li, T. Mhlanga, M. Escuti, and A. Forbes, ``Generating and measuring nondiffracting vector Bessel beams," \ol {\bf38}(17), 3429-3432 (2013).
\bibitem{sorter}A. Dudley, T. Mhlanga, M. Lavery, A. McDonald, F. S. Roux, M. J. Padgett, and A. Forbes, ``Efficient sorting of Bessel beams, '' \opex {\bf21}(1), 165-171 (2013).
\bibitem{mazilu2} A. Mourka, M. Mazilu, E. M. Wright, and K. Dholakia  ``Modal characterization using principal component analysis: application to Bessel, higher-order Gaussian beams and their superpositions,'' Scientific Reports {\bf 3}, 1422 (2013).
\bibitem{mazilu3} M. Mazilu, A. Mourka, T. Vettenburg, E. M. Wright, and K. Dholakia  ``Simultaneous determination of the constituent azimuthal and radial mode indices for light fields possessing orbital angular momentum,'' Appl. Phys. Lett. {\bf 100}, 231115 (2012).
\bibitem{obstruction}I. Litvin, M. McLaren, and A. Forbes, ``A conical wave approach to calculating Bessel-Gauss beam reconstruction after complex obstacles,'' \oc {\bf282}, 1078-1082 (2009).
\bibitem{obstruction2}Z. Bouchal, J.Wanger, M. Chulpl, ``Self-reconstruction of a distorted nondiffracting beam,'' \oc {\bf151}, 207-211 (1998).
\bibitem{OAM1}I. A. Litvin, A. Dudley, F. S. Roux, and A. Forbes, ``Azimuthal decomposition with digital holograms,'' \opex {\bf20}(10), 10996-11004 (2012).
\bibitem{SR}A. Janssen, S. van Haver, P. Dirksen, J. Braat,``Zernike representation and strehl ratio of optical systems with numerical aperture,''J. Mod. Opt. {\bf 55}(7), 1127-1157 (2008).
\bibitem{turbulence1}J. Ou, Y. Jiang, J. Zhang, H. Tang, Y. He, S. Wang, J. Liao, ``Spreading of spiral spectrum of Bessel-Gaussian beam in non-Kolmogorov turbulence,'' \oc {\bf318}, 95-99 (2014).
\bibitem{turbulence2}W. Nelson, J. P. Palastro, C. C. Davis, and P. Sprangl, ``Propagation of Bessel and Airy beams through atmospheric turbulence,'' \josaa {\bf31}(3), 603-609 (2014).


\end{thebibliography}
\end{document}